\documentclass[12pt]{article}
\usepackage{latexsym}
\usepackage{graphicx}
\newcommand{\beq}{\begin{equation}}
\newcommand{\eeq}[1]{\label{#1}\end{equation}}
\newcommand{\bea}{\begin{eqnarray}}
\newcommand{\eea}[1]{\label{#1}\end{eqnarray}}

\begin{document}
\begin{flushright}
\hfill{MCTP-02-43}\\
\hfill{hep-th/0208093 }
\end{flushright}

\centerline{\bf Comment on time-variation of fundamental constants}
 \vspace*{0.37truein} \centerline{\footnotesize M.
J.  Duff\footnote{mduff@umich.edu. }} \vspace*{0.015truein}
\centerline{\footnotesize\it Michigan Center for Theoretical Physics}
\baselineskip=10pt
\centerline{\footnotesize\it Randall Laboratory, Department of Physics,
University of Michigan}
\centerline{\footnotesize\it Ann Arbor, MI 48109--1120, USA}
\bigskip


\vspace{20pt}

\abstract{The possible time variation of dimensionless fundamental
constants of nature, such as the fine-structure constant $\alpha$, is a
legitimate subject of physical enquiry.  By contrast, the time
variation of dimensional constants, such as $\hbar$, $c$, $G$, $e$,
$k$\ldots, which are merely human constructs whose number and values
differ from one choice of units to the next, has no operational
meaning.  To illustrate this, we refute a recent claim of Davies et
al that black holes can discriminate between two contending theories of
varying $\alpha$, one with varying $c$ and the other with varying $e$.
In Appendix A we respond to criticisms by P. Davies and two {\it Nature}
referees.  In Appendix B we respond to remarks by Magueijo and by 
T. Davis. In Appendix C we critique recent claims by Copi, A. Davis 
and Krauss to have placed constraints on $\Delta G/G$.} In Appendix D we provide extracts of a lecture by Dirac, of which we have only recently  become aware, which includes the comment ``Talking about whether a thing is constant or not  does not have any absolute meaning 
unless that quantity is dimensionless ''.


\section{\bf Black holes and varying constants}

The claim \cite{Webb} that the fine-structure
constant, $\alpha$-the measure of the strength of the electromagnetic
interaction between photons and electrons-is slowly increasing over cosmological
time scales has refuelled an old debate about varying fundamental
constants of nature. In our opinion \cite{Duff:2001ba}, however, this debate
has been marred by a failure to distinguish between {\it dimensionless}
constants such as $\alpha$, which may indeed be fundamental, and {\it
dimensional} constants such as the speed of light $c$, the charge
on the electron $e$, Planck's constant $\hbar$, Newton's constant $G$, Boltzmann's
constant $k$ etc, which are merely human constructs whose number and
values differ
from one choice of units to the next and which have no intrinsic physical
significance. An example of this
confusion is provided
by a recent paper \cite{Davies:zz}, where it is claimed that ``As
$\alpha=e^{2}/{\hbar}c$, this would call into question which of these
fundamental quantities are truly constant''.  By consideration of
black hole thermodynamics, the authors conclude that theories with
decreasing $c$ are different from (and may be favored over) those with 
increasing $e$.  Here we argue that this claim is operationally 
meaningless, in the sense that no experiment could tell the 
difference, and we replace it by a meaningful one involving just 
dimensionless parameters.

Any theory of gravitation and elementary particles is characterized by
a set of dimensionless parameters such as coupling constants
$\alpha_{i}$ (of which the fine-structure constant is an example),
mixing angles $\theta_{i}$ and mass ratios $\mu_{i}$.  To be concrete,
we may take $m_{i}{}^{2}=\mu_{i}{}^{2}\hbar c/G$ where $m_{i}$ is the
mass of the i'th particle.  The Standard Model has 19 such parameters,
but it is hoped that some future unified theory might reduce this
number.  By contrast, the number and values of dimensional constants,
such as $\hbar$, $c$, $G$, $e$, $k$ etc, are quite arbitrary human
conventions.  Their job is merely to convert from one system of units
to another.  Moreover, the more units you introduce, the more such
conversion factors you need \cite{Duff:2001ba}.

The authors of \cite{Davies:zz} point out that the entropy $S$ of a
non-rotating black hole with charge $Q$ and mass $M$ is given by
\begin{equation}
S=\frac{k\pi G}{\hbar c}[M+\sqrt{M^{2}-Q^{2}/G}]^{2}
\label{entropy}
\end{equation}
They note that decreasing $c$ increases $S$ but increasing $e$, and hence
$Q$, decreases $S$.  It is then claimed, erroneously in our view, that
black holes can discriminate between two contending theories of
varying $\alpha$, one with varying $c$ and the other with varying $e$.

Let us define the dimensionless parameters $s$, $\mu$ and $q$ by
$S=sk\pi$, $M^{2}=\mu^{2}\hbar c/G$ and $Q^{2}=q^{2}\hbar c$.  The 
mass ratio $\mu$ 
will depend on the fundamental dimensionless parameters of the theory 
$\alpha_{i}$, $\theta_{i}$ and $\mu_{i}$, but the details need not 
concern us here. In the
Appendix we shall give a thought-experimental definition of $s$, $\mu$
and $q$ that avoids all mention of the unit-dependent quantities $G$,
$c$, $\hbar$, and $e$ and which is valid whether or not they are
changing in time.  
Shorn of all its irrelevant unit dependence, therefore,
the entropy is given by
\begin{equation}
s= [\mu+\sqrt{\mu^{2}-q^{2}}]^{2}
\label{entropy1}
\end{equation}
If we use the fact that the charge is quantized in units of $e$, namely
$Q=ne$ with $n$ an integer, then $q^{2}=n^{2}\alpha$, but we prefer not
to mix up macroscopic and microscopic quantities in (\ref{entropy}).  So 
the correct conclusion is that such black holes might\footnote{Since 
the purpose of this paper is to
critique the whole school of thought, of which Ref. \cite{Davies:zz}
is but an example, that believes time-dependent dimensional parameters
have operational significance, we will not divert attention by getting
into the question of whether black hole thermodynamics provides a good laboratory
for testing time variation of dimensionless constants.  Suffice it to
say that that, in our view, the only sensible context in which to
discuss time varying constants of nature is in theories where they are
given by moduli (i.e. vacuum-expectation-values of scalar fields).
The black hole entropy would then be expressed in terms of these moduli
\cite{Gibbons} whose time-dependence would have to be determined
consistently by the field equations.  Moreover, one would have to take
into account both the entropy of the black hole and its environment.} discriminate
between contending theories with different variations of $\mu$ and
$q$.

The unit dependence of the claim in \cite{Davies:zz} that black holes
can discriminate between varying $c$ and varying $e$ is now evident.
In Planck units \cite{Planck,Duff:2001ba}
\begin{equation}
\hbar=c=G=1~~~~~~~e^{2}=\alpha~~~~~~~M^{2}�=\mu^{2}�
\label{Planck1}
\end{equation}
In Stoney units \cite{Stoney,Duff:2001ba}
\begin{equation}
c=e=G=1~~~~~~~{\hbar}=1/\alpha~~~~~~~M^{2}�=\mu^{2}�/\alpha
\label{Stoney1}
\end{equation}
In Schrodinger units (see Appendix)
\begin{equation}
\hbar=e=G=1~~~~~~~c=1/\alpha~~~~~~~M^{2}�=\mu^{2}�/\alpha
\label{Schrodinger1}
\end{equation}
In all three units (and indeed in {\it any}\footnote{Units in which 
$G$ varies are discussed in the Appendix.} units), the dimensionless 
entropy ratio $s$ is the same as given by (\ref{entropy}).  To 
reiterate: assigning a change in $\alpha$ to a change in $e$ (Planck) 
or a change in $\hbar$ (Stoney) or a change in $c$ (Schrodinger) is 
entirely a matter of units, not physics.  Just as no experiment can 
determine that MKS units are superior to CGS units, or that degrees 
Fahrenheit are superior to degrees Centigrade, so no experiment can 
determine that changing $c$ is superior to changing $e$, contrary to 
the main claim of Davies et al \cite{Davies:zz}.

In summary, it is operationally meaningless \cite{Duff:2001ba} and
confusing to talk about time variation of arbitrary unit-dependent
constants whose only role is to act as conversion factors.  For
example, aside from saying that $c$ is finite, the statement that
$c=3\times 10^{8}~m/s$, has no more content than saying how we convert
from one human construct (the meter) to another (the second).  Asking
whether $c$ has varied over cosmic history (a question unfortunately
appearing on the front page of the {\it New York Times} \cite{Glanz}, 
in {\it Physics
World} \cite{PW1}\footnote{To its credit, {\it Physics World} published
the opposing
view \cite{PW2}.}, in {\it New Scientist} \cite{NS1,NS2,NS3}, in {\it Nature} \cite{Davies:zz}
and on {\it CNN} \cite{CNN}) is like asking whether the number of liters
to the
gallon has varied.

 \section
{Acknowledgements} I am grateful to Thomas Dent for drawing my
attention to Ref. \cite{Davies:zz} and for valuable discussions, to
Samir Mathur and Parthasarathi Mitra for suggesting improvements and 
to Joao Magueijo for correspondence. This research
supported in part by DOE Grant DE-FG02-95ER40899.

\appendix

\section{Response to criticisms}

Since, after receiving a comment from Davies and two referee reports,
{\it Nature} rejected (a shorter version of) this paper without
the opportunity to rebut the criticisms, we do so in this
Appendix.  In the opinion of the present author, these three 
commentaries contain many of the most common misconceptions regarding 
fundamental constants.  For example:

{\it Referee 1}: It is true that if the fundamental ``constants'' $h$, 
$c$, $G$, $k$...are truly
constant, then they do indeed only act as conversion factors and can
e.g. be
set equal to unity. However, when they are postulated (or discovered
experimentally to vary) in time, then we have to take into account that
varying one or the other of these constants can have significant
consequences for physics\footnote{Similar objections were raised by 
Moffat \cite{Moffat}.}.  Thus, varying the charge $e$ will have 
different experimentally testable consequences than varying either $h$ 
or $c$.

{\it Response}: To elaborate more on our refutation of this common
fallacy, 
let us define Planck \cite{Planck,Duff:2001ba} length, mass, time and
charge by
\[
L_{P}{}^{2}=G\hbar /c^3
\]
\[
 M_{P}^{2}=\hbar c/G
 \]
 \[
 T_{P}{}^{2}=G\hbar/c^5
 \]
 \begin{equation}
 Q_{P}{}^{2}=\hbar c
\label{Planck}
\end{equation}
Note that these are independent of $e$, since Planck was not
immediately concerned with electrodynamics in this context.  Now
define Stoney \cite{Stoney,Duff:2001ba} length, mass, time and charge by
   \[
L_{S}{}^{2}=Ge^{2}/c^4=\alpha L_{P}{}^{2}
\]
\[
M_{S}^{2}=e^{2}/G=\alpha M_{P}{}^{2}
\]
\[
T_{S}{}^{2}=Ge^{2}/c^6=\alpha T_{P}{}^{2}
\]
\begin{equation}
Q_{S}{}^{2}=e^{2}=\alpha Q_{P}{}^{2}
\label{Stoney}
\end{equation}
Note that these are independent of $\hbar$ since Stoney knew nothing
of quantum mechanics.  They are obtained from Planck units by the
replacement $\hbar \rightarrow \alpha \hbar$.  To complete the trio we
need units that take $e$ and $\hbar$ into account but are independent
of $c$.  It seems appropriate, therefore, to call these Schrodinger
length, mass, time and charge, defined by
\[
L_{\psi}{}^{2}=G\hbar^{4}/e^6=L_{P}{}^{2}/\alpha^{3}
\]
\[
M_{\psi}^{2}=e^{2}/G=\alpha M_{P}{}^{2}
\]
\[
T_{\psi}{}^{2}=G\hbar^{6}/e^{10}=T_{P}{}^{2}/\alpha^{5}
\]
\begin{equation}
Q_{\psi}{}^{2}=e^{2}=\alpha Q_{P}{}^{2}
\label{Schrodinger}
\end{equation}
They are obtained from Planck units by making the replacement
$c \rightarrow \alpha c$.

We now ask the question: can we give a thought-experimental (as
opposed to purely mathematical) meaning to the above length, mass,
time and charge units that is valid whether or not quantities are 
changing in time?  Interestingly, the ERNBH (extreme
Reissner-Nordstrom black hole) solution provides the answer.  A
non-rotating black hole with charge $Q$ and mass $M$ has Schwarzschild
radius
\begin{equation}
R_{S}=\frac{GM}{c^{2}}+\sqrt{\frac{G^{2}M^{2}}{c^{4}}
-\frac{GQ^{2}}{c^{4}}}
\end{equation}
and Compton wavelength
\begin{equation}
R_{C}= \frac{h}{Mc}.
\end{equation}
In the extreme case, moreover, we have
\begin{equation}
R_{S}=\frac{GM}{c^{2}}~~~~~~Q^{2}=GM^{2}
\end{equation}
$L_{P}$, $M_{P}$, $T_{P}$ and $Q_{P}$ may now be thought-experimentally
defined 
without reference to any fundamental constants as
the Schwarzschild radius, mass, characteristic time and charge of an
ERNBH whose Schwarzschild radius equals its Compton wavelength divided
by $2\pi$.  Thus $s$, $\mu$ and $q$ count the number of times $S$, $M$
and 
$Q$ exceed the entropy, mass and charge of such an black hole.  
Similarly, $L_{S}$, $M_{S}$, $T_{S}$ and $Q_{S}$ are the corresponding 
quantities for an ERNBH whose charge is the charge on the electron.  
$L_{\psi}$, $M_{\psi}$, $T_{\psi}$ and $Q_{\psi}$ also admit a 
thought-experimental definition which we defer until the introduction 
of Bohr units below.

So even if dimensionless constants are
changing in time, nothing stops us from using Planck units with
$c=\hbar=1$ and time varying $e$, Stoney units with $c=e=1$ and
time varying $\hbar$ or Schrodinger units with $\hbar=e=1$ and time
varying $c$.

{\it Referee 1}: It is conceivable that varying the charge $e$ could
lead to a theory that
somehow could be re-written as a theory in which $e$ is kept fixed and c is
varied, but this would lead to a strange and very complicated revision of
all of physics.

{\it Response}: On the contrary, it is nothing more than switching from
Planck units to Schrodinger units.

{\it Referee 1}: He has already published his views on this issue
in: M. J. Duff, L. B. Okun and G. Veneziano, JHEP 0203, 023 (2002). It
is to
be noted that the other two authors of this article do not appear to agree
with Duff that it is ``operationally meaningless'' to vary dimensional
constants.

{\it Response}: According to this logic, the present paper would have to
be 
rejected whichever of the three mutually contradictory views in 
\cite{Duff:2001ba} were being put forward.  I think it should be 
judged on its own merits.

{\it Referee 1}: Perhaps, Duff would wish to clarify his position on the
issue of how the rest of physics is affected by a possible variation of 
dimensional constants. The issue of how physics would be affected by the
experimental discovery
that ``constants'', such as c, h, G... vary in time is clearly of fundamental
importance. 

{\it Response}: I believe my position is clear: physics is about 
dimensionless constants and is completely unaffected by the choice of 
units, which has no fundamental importance.

{\it Referee 2}: Physics without reference to dimensional 
quantities is unfortunately not a possibility.  Curiously this fact 
only shocks the author with reference to a changing $c$.

{\it Response}: I beg to differ.  Dimensional quantities may sometimes 
be useful, but from an empirical point of view, experiments measure only 
dimensionless quantities.  From a theoretical point of view, moreover,
any 
theory may be cast into a form in which no dimensional quantities ever 
appear either in the equations themselves or in their solutions (such 
as vacuum expectation values of scalar fields).  So the issue of 
whether dimensional parameters vary in time need never arise.  
Moreover, it should be clear from the text that my objections apply to 
all dimensional constants, not just $c$.

{\it Referee 2}:  When one says that the speed of light is color
dependent (as in the case
for deformed dispersion relations), or that the speed of light
and gravity vary with respect to each other (as for some bimetric
theories 
of gravity), one makes dimensionless statements.

{\it Response}: Although all fundamental constants are dimensionless, 
the converse is not true.  For example, the ratio of the Earth's 
radius and the Sun's radius is an accident of nature with no 
fundamental significance.  Moreover, not all dimensionless quantities 
are unit independent.  For example $\delta c/c$ is zero in Planck and 
Stoney units but non-zero in Schrodinger units.

{\it Referee 2}: Also Lorentz invariance has an operational sense and
some aspects of the constancy of $c$ are directly related to it.

{\it Response}: Let us suppose that we have a generally covariant and 
locally Lorentz invariant theory of gravity with scalar fields, and that
time varying $\alpha$ is implemented by a time-dependent scalar field 
solution. 
This background will not exhibit global Lorentz invariance, but this is
no 
different than a time-dependent Friedman-Robertson-Walker cosmology 
which is not Lorentz-invariant either. Alternatively, we might 
imagine a phase transition from one Lorentz-invariant vacuum to another 
in which the dimensionless constants, such as $\alpha$, change abruptly.
Whatever 
the symmetries, they 
will be the same whether we use varying $c$ units or some other units. 
Moreover, none of this conflicts with Einstein's general 
covariance, contrary to certain claims in the literature and in the
media. 

{\it Referee 2}: The last phrase of the paper is wrong 
(the same argument could be applied to variations in e or entropy, 
after all).

{\it Response}: The last sentence could indeed be applied to any 
other conversion factor but is nevertheless correct.

{\it Davies}: Where 
we differ substantially from Duff, and where it seems clear he is 
wrong, is in his claim that theories in which dimensional constants 
vary with time ``is operationally meaningless.'' Such theories have 
existed in the literature, and specific observational tests been 
suggested and carried out, at least since Dirac's theory of varying $G$.

{\it Response}: I agree that Davies et al are the latest in a long line of
authors making such claims, but Dirac was not one of them.  In his 
seminal paper \cite{Dirac} he says: ``The fundamental constants of 
physics, such as $c$ the velocity of light, $h$ the Planck constant, 
$e$ the charge and $m_{e}$ the mass of the electron, and so on, 
provide for us a set of absolute units for measurement of distance, 
time, mass, etc.  There are, however, more of these constants than are 
necessary for this purpose, with the result that certain dimensionless 
numbers can be constructed from them.'' The phrase ``more of these 
constants than are necessary'' is crucial.  Those who insist on 
counting the dimensional constants in a theory as well as the 
dimensionless ones will always have more unknowns than equations.  
This redundancy is nothing but the freedom to change units without 
changing the physics.  In Einstein-Maxwell-Dirac theory, for example, 
one could imagine units in which (at least) five dimensional 
constants, are changing in time: $G$, $e$, $m_{e}$, $c$, 
$\hbar$\ldots, but only two dimensionless combinations are necessary: 
$\mu_{e}{}^{2}=Gm_{e}{}^{2}/\hbar c$ and $\alpha=e^{2}/\hbar c$.

Dirac then notes that the {\it dimensionless} ratio of electromagnetic and
gravitational forces $e^{2}/Gm_{e}^{2}$ is roughly the same order of
magnitude as the {\it dimensionless} ratio of the present age of the
universe 
$t$ and the atomic unit of time $e^{2}/m_{e}c^{3}$.  He makes it clear
that 
equating these two numbers leads to a time-varying $G \sim t^{-1}$ only
in these 
``atomic units''. To be explicit, let us define Dirac \cite{Dirac}
length, mass,
time and charge by
\[
L_{D}{}^{2}=e^{4}/m_{e}{}^{2}c^4=L_{S}^{2}\alpha/\mu_{e}{}^{2}
\]
\[
 M_{D}^{2}=m_{e}{}^{2}=M_{S}^{2}\mu_{e}{}^{2}/\alpha
 \]
 \[
 T_{D}{}^{2}=e^{4}/m_{e}{}^{2}c^6=T_{S}{}^{2}\alpha/\mu_{e}{}^{2}
 \]
 \begin{equation}
 \label{Dirac}
 Q_{D}{}^{2}=e^{2}=Q_{S}^{2}
\end{equation} 
Note that these units are independent of $G$ and $\hbar$.  They are
obtained from
Stoney units by the replacement $G \rightarrow G\alpha
/\mu_{e}{}^{2}$.  In Dirac units
\begin{equation}
c=e=m_{e}=1~~~~~\hbar=1/\alpha~~~~~G=\mu_{e}{}^{2}/\alpha
~~~~M^{2}=\mu^{2}/\mu_{e}{}^{2}
\label{Dirac1}
\end{equation}
Once again, the entropy is the same as given by (\ref{entropy}).
So there is no such thing as a varying $G$ {\it theory}, only 
varying $G$ {\it units}. This is familiar from string theory \cite{GSW}
where the 
string tension $T$ is related to $G$ via dilaton and moduli fields 
which may possibly vary in space and time.  In Einstein units, $G$ is
fixed 
while $T$ may vary, whereas in string units $T$ is fixed while $G$ may 
vary.
 
{\it Davies}: Some
theories of fundamental physics, e.g. the Hoyle-Narlikar theory of
gravitation, were explicitly designed to incorporate an additional gauge
freedom (in that case, conformal invariance) to enable one to transform at
will between different systems of units, without changing the physics,
whilst including cosmological time variations of constants. 

{\it Response}: The freedom to choose MKS units, say, over CGS units 
requires no symmetry of the fundamental theory but is merely one of
human 
convention.  The same is true of choosing changing $c$ units
 over changing $e$ units. 
  
   {\it Davies}: Several varying speed of light and 
  varying electric charge theories have been published, and explicit 
  observational predictions made.  See, for example, J.  Magueijo, Phys.
  Rev.  D.  62, 103521 (2000), and ``Is it e or is it c?  Experimental 
  tests of varying alpha'' by J.  Magueijo, J.D.  Barrow and H.B.  
  Sandvik, Phys.  Rev.  D, in the press (available online at arXiv: 
  astro-ph/0202374v1, 20 February 2002).

{\it Response}: This seems a curious choice of authors to back up
Davies' argument.
In his paper with Albrecht \cite{Albrecht}, Magueijo says `` Our
conclusion that 
physical experiments are only sensitive to dimensionless combinations 
of dimensional constants is hardly a new one. This idea has been often 
stressed by Dicke (eg. \cite{Dicke}), and we believe this is not 
controversial.'' Majueijo, Barrow and Sandvik \cite{Barrow} say 
``Undoubtedly, in the sense of
\cite{Duff:2001ba}, one has to make an operationally `meaningless'
choice of which dimensional constant is to become a dynamical
variable.'' These papers thus fall into a category whose authors are well
aware that there is no experimental way of distinguishing varying $e$
from varying $c$, but nevertheless choose to label genuinely
physically inequivalent theories by the names ``varying $c$'' and
``varying $e$'' merely  because they find one set of units more convenient
than the other.  I might criticize these papers for being 
confusing\footnote{While recognizing that time variation of dimensional 
quantities lacks operational definition, Carlip and Vaidya \cite{Carlip} 
nevertheless try to salvage from subjectivity the notion of changing
$e$, for 
example, by saying ``suitable variation of all dimensionless 
parameters that depend on $e$''.  But this is equally subjective: 
nature provides us with dimensionless parameters and humans decide 
where to put the $c$'s etc.  For example, which of the 19 parameters 
of the Standard Model depend on $c$?  It is entirely up to you!}, but 
not wrong.  They provide cold comfort for Davies et al who claim that 
varying $c$ and varying $e$ are experimentally distinguishable.

{\it Davies}: The speed of light is more than an electrodynamic
parameter: it
describes the causal structure of spacetime, and as such is relevant to
all of
physics (for example, the weak and strong interactions), not just
electrodynamics. 

{\it Response}: What is relevant for the strong, weak and 
electromagnetic interactions is the special theory of relativity, i.e
invariance under the Poincare group of spacetime transformations. 
The mathematics of the Poincare group ($x'{}^{\mu}=
\Lambda^{\mu}{}_{\nu}x^{\nu}+ a^{\mu}$)
can get along just fine without $c$.

{\it Davies}: A variation of $c$ cannot be mimicked in all such respects
by a
change in $e$.  More obviously, one can imagine measuring the speed of
light in the
laboratory tomorrow and obtaining a different value from today. That is
clearly operationally meaningful.

{\it Response}: This common fallacy can be eliminated by thinking carefully
about how one would attempt to measure $c$ in a world in which
dimensionless constants such as $\alpha$ and $\mu$ are changing in
time.  First take a ruler with notches one Planck length apart and a
clock with ticks one Planck time apart.  Next measure the speed of
light in vacuum\footnote{If the experiment is performed in a medium, 
or a time-dependent gravitational field, one would have to factor out
the 
effects of the refractive index, or $\sqrt{g_{xx}/g_{tt}}$.  After all, 
light slows down when passing through a piece of glass, but no-one is 
suggesting that this produces an increase in $\alpha$.} by counting 
how many notches light travels in between ticks.  You will find the 
answer $c=1$.  You may repeat the experiment till the cows come home 
and you will always find $c=1$!  Repeat the experiment using Stoney 
length and Stoney time, and again you will find $c=1$.  But if the 
notches on your ruler are one Schrodinger length apart and the ticks 
on your clock one Schrodinger time apart, you will find $c=1/\alpha$ 
and $c$ will now have the same time dependence as $1/\alpha$.  Once 
again we see that the time dependence of $c$ is entirely 
unit-dependent.  Similar remarks apply to the measurement of any other 
dimensional quantity.

For the sake of completeness, let us also define Bohr length, mass, 
time and charge, which have an obvious atomic definition as the Bohr
radius 
etc:
\[
L_{B}{}^{2}=\hbar^{4}/m_{e}{}^{2}e^4=L_{\psi}^{2}\alpha/\mu_{e}^{2}
\]
\[
M_{B}^{2}=m_{e}{}^{2}=M_{\psi}^{2}\mu_{e}^{2}/\alpha
 \]
 \[
T_{B}{}^{2}=\hbar^{6}/m_{e}{}^{2}e^8=T_{\psi}{}^{2}\alpha/\mu_{e}^{2}
 \]
\begin{equation}
Q_{B}{}^{2}=e^{2}=Q_{\psi}^{2}
\end{equation}
Note that these units are independent 
of $G$ and $c $.  They are obtained from Schrodinger units by the
replacement 
$G \rightarrow G \alpha/\mu_{e}{}^{2}$. In Bohr units
\begin{equation}
\hbar=e=m_{e}=1~~~~~c=1/\alpha~~~~~G=\mu_{e}{}^{2}/\alpha~~~~
~~~~M^{2}=\mu^{2}/\mu_{e}{}^{2}
\end{equation}
So a thought experimental definition of Schrodinger $L_{\psi}{}^{2}$ is
the Bohr 
$L_{B}{}^{2}$ scaled down by Dirac's large number (the ratio of 
electromagnetic to gravitational forces $e^{2}/Gm_{e}^{2}$) with 
similar definitions for $M_{\psi}^{-2}$ and $T_{\psi}^{2}$.

Measuring the speed of light with a ruler whose notches are one 
Bohr length apart and a clock whose ticks are one Bohr time apart will 
again result in $c=1/\alpha$.  As discussed in \cite{Albrecht}, Bohr 
units are used when measuring $c$ using an atomic clock, which is most 
sensitive to a variation of $\alpha$.  A pendulum clock, on the other 
hand, is more sensitive to the variation of $\mu_{i}$.  So when you 
think you are measuring a dimensional quantity, you are really 
measuring dimensionless ones.

{\it Davies}: So this is an issue of semantics and mathematical
elegance, not
science.

{\it Response}: The failure to tell the difference between changing
units and 
changing physics is more than just semantics. It brings to mind the 
old lady who, when asked by the TV interviewer whether she believed in  
global warming, responded: ``If you ask me, it's all this changing 
from Fahrenheit to Centigrade that's causing it''.

\section{Response to remarks on the present paper}

\subsection{Remarks by Magueijo} 
 
In the abstract of a recent review article \cite{Review}, Magueijo 
writes:  ``We start by discussing the
physical meaning of a varying $c$, dispelling the myth that the
constancy of $c$ is a matter of logical consistency''. The following 
statements appear in the text.

{\it Magueijo}: In discussing the physical meaning of a varying speed of light,
I'm afraid that Eddington's religious fervor is still with
us \cite{Ellis,Duff}. ``To vary the speed of light is
self-contradictory'' has now been transmuted into ``asking whether
$c$ has varied over cosmic history is like asking whether the
number of liters to the gallon has varied''
\cite{Duff}. The implication is that the constancy of
the speed of light is a logical necessity, a definition that could
not have been otherwise. This has to be naive. For centuries the
constancy of the speed of light played no role in physics, and
presumably physics did not start being logically consistent in
1905. Furthermore, the postulate of the constancy of $c$ in
special relativity was prompted by experiments (including those
leading to Maxwell's theory) rather than issues of consistency.
History alone suggests that the constancy (or otherwise) of the
speed of light has to be more than a self-evident necessity.

{\it Response}: In fact my remark implies no such logical necessity. 
It merely means
that the variation or not of dimensional numbers like c (as opposed to
dimensionless numbers like the fine-structure constant) is a matter
human convention, just as the variation or not in the number of liters
to a gallon is a matter of human convention. In neither case is it
something to be determined by experiment but rather by one's choice of
units.� So there is no such thing as a varying $c$ `theory' only 
varying $c$ `units'.� For example, in units where time is measured in years and
distance in light-years, $c=1$ for ever and ever, whatever your theory!

As a matter of fact, the number of liters per gallon varies as 
one crosses the Atlantic. Similarly, as discussed in 
\cite{Duff:2001ba}, in 1983 the Conference Generale des Poids 
et Mesures changed the number of meters per second, i.e the value of c.  Relativity 
survived intact!

\subsection{Remarks by T. Davis}

In a recent paper \cite{Davis}, one of the authors of the black hole thermodynamics 
paper, T. Davis, responds to our criticisms. With reference to our equation 
(\ref{entropy1}), the following appears in the text:

{\it T. Davis}: Arguments from quantum theory suggest that it is more natural to 
expect that $\mu$ would remain constant than $M$.  Under this assumption 
Eq.~(\ref{entropy1}) suggests that any increase in $\alpha$ would 
violate the second law of thermodynamics, independent of which of 
$e$, $c$ or $\hbar$ varies.  This seems to be in contradiction to our 
previous result in which an increase in $e$ decreased $s$ but 
a decrease in $c$ or $\hbar$ increased $s$.  However, in our 
initial formulation we had assumed that $M$ remained constant whereas 
here we are assuming $\mu$ remains constant.  If we allow $\mu$ to vary such 
that $M$ remains constant the result for black hole entropy is unchanged 
from the previous version.

{\it Response}: The variation or not of the dimensionless quantity $\mu$ 
is operationally meaningful, but once again, the question of whether or not the 
dimensional quantity M varies is a matter of human convention,
not quantum theory. In Planck units (\ref{Planck1}), for example, 
variation of $\mu$ implies variation of $M$ while in Stoney units 
(\ref{Stoney1}) or Schrodinger units (\ref{Schrodinger1}), it does not. 
The black hole entropy (\ref{entropy1}) and the second law of 
thermodynamics do not give a fig which units are chosen.

\section{Comments on claims to have placed constraints on $\Delta G/G$}

A recent paper by Copi, A. Davis and Krauss \cite {Krauss} claims to use 
astrophysical data to place constraints on the time variation  of 
Newton's constant, $\Delta G/G$. Here we reiterate the point made in 
Appendix A that dimensionless 
ratios such as $\Delta G/G$, $\Delta e/e$ and $\Delta c/c$ are every 
bit as unit-dependent as their dimensional counterparts $\Delta G$, 
$\Delta e$ and $\Delta c$. An obvious example is again provided by 
units in which time is measured in years and distance in light-years.
Here $c=1$ and $\Delta c/c$=0, whatever your theory. Similar remarks 
apply to $\Delta G/G$. As discussed in Appendix A, it is guaranteed to 
vanish in Planck units (\ref{Planck1}), for example, but might vary in Dirac 
units (\ref{Dirac1}). By contrast, $\Delta \alpha/\alpha$ is 
unit-independent.

The idea of varying $G$ is frequently attributed to the papers of Dirac 
\cite{Dirac} and \cite{Dicke} Dicke, 
but as discussed in Appendix A, a reading of these papers reveals that 
both authors were aware that it is only 
dimensionless numbers such as $\mu_{e}$  and $\alpha$ that are operationally 
meaningful. The Standard Model coupled to gravity with a cosmological 
constant has 20 such parameters. It is the variation of these quantities 
that may be contrained by the astrophysical data presented in 
\cite{Krauss}, not $\Delta G$ nor even $\Delta G/G$.

\section{Extracts from a recently discovered Dirac lecture}

It was not until 2016 that I became aware of two lectures by Paul Dirac on Dimensionless Physical Constants and his ``Large Number Hypothesis"

https://www.youtube.com/watch?v=-o8mUyq\_Wwg

https://www.youtube.com/watch?v=P174LmmQYy4

Here are some extracts:

{\it Dirac:} I would like to think about the constants of nature. Most of these have dimensions and their numerical value depends on what system of units you are using. Such numerical values of course are not of any general interest. However, one can construct some constants of nature which are dimensionless and these will have the same numerical value whatever units one uses. They have a numerical value which is thus provided by nature quite independent of man-made units. 

One of these numbers is the reciprocal of the fine structure constant 
\[
\hbar c/e^2 \sim 137. 
\]
There is another one of these dimensionless numbers I would like to call to your attention. The most natural one is the ratio of the mass of the proton to the mass of the electron 
\[
m_p/m_e \sim 1840.
\]
Another provided by nature is the ratio of the electric force and the gravitational force between a proton and an electron 
\[
e^2/Gm_pm_e \sim 10^{39},
\]
a very large number.  

The age of the universe measured in atomic units of time $e^2/m_ec^3$ is another large dimensionless number rather close to the one we had before, namely $10^{39}$. I believe this is not a coincidence. Let us accept that there is a connection between these two numbers. Then at least one of the quantities $e, G, m_e, m_p,c$, which are usually considered as constants, should be varying in time. 

Which one varies in time? That is rather a meaningless question.

Talking about whether a thing is constant or not  does not have any absolute meaning unless that quantity is dimensionless.

For example, it doesn't have any absolute meaning to talk about G varying because G has dimensions.

{\it Response}

I invite the Editor of {\it Nature} to comment.


\end{document}